\documentclass[aps,prl,reprint,superscriptaddress,twocolumn,notitlepage]{revtex4-2}
\usepackage{amsmath}
\usepackage{amsfonts}
\usepackage{amssymb}
\usepackage{graphicx}
\usepackage{multirow}
\usepackage{chngcntr}
\usepackage{bm}
\usepackage[T1]{fontenc} 
\usepackage{ulem}
\usepackage[usenames,dvipsnames]{xcolor}

\usepackage{natbib}
\usepackage[final=true]{hyperref}
\hypersetup{
   colorlinks=true,
    linkcolor=Maroon,
   filecolor=magenta,      
    urlcolor=cyan,
    citecolor=PineGreen,    
}

\newcommand{\m}{\mathbf}

\newcommand{\be}{\begin{eqnarray}}
\newcommand{\ee}{\end{eqnarray}}
\newcommand{\nn}{\nonumber}

\newcommand{\sect}[1]{\textit{#1}---}

\begin{document}

\title{
Subcritical bifurcation and on-off bistability in ballistic polariton condensates}

\author{Oleg I. Utesov}

\email{utiosov@gmail.com}
\thanks{These authors contributed equally to this work.}
\affiliation{Department of Physics, Korea Advanced Institute of Science and Technology (KAIST), Daejeon 34141, Republic of Korea}
\affiliation{Center for Theoretical Physics of Complex Systems, Institute for Basic Science (IBS), Daejeon 34126, Republic of Korea}

\author{Soohong Choi}
\thanks{These authors contributed equally to this work.}
\affiliation{Department of Physics, Korea Advanced Institute of Science and Technology (KAIST), Daejeon 34141, Republic of Korea}

\author{Pavel Kozhevin}
\affiliation{Department of Physics, St. Petersburg State University, St. Petersburg 199034, Russia}
\affiliation{Abrikosov Center for Theoretical Physics, Moscow Institute of Physics and Technology, Dolgoprudnyi, Moscow Region 141701, Russia}

\author{Min Park}\affiliation{Department of Physics, Korea Advanced Institute of Science and Technology (KAIST), Daejeon 34141, Republic of Korea}

\author{Daegwang Choi}\affiliation{Department of Physics, Korea Advanced Institute of Science and Technology (KAIST), Daejeon 34141, Republic of Korea}

\author{Hyungdo Lee}
\affiliation{Department of Physics, Korea Advanced Institute of Science and Technology (KAIST), Daejeon 34141, Republic of Korea}

\author{Alexey N. Osipov}\affiliation{Department of Physics, ITMO University, St. Petersburg 197101, Russia}

\author{Alexey V. Yulin}\affiliation{Department of Physics, ITMO University, St. Petersburg 197101, Russia}

\author{Se Kwon Kim}
\affiliation{Department of Physics, Korea Advanced Institute of Science and Technology (KAIST), Daejeon 34141, Republic of Korea}
\affiliation{Graduate School of Quantum Science and Technology, Korea Advanced Institute of Science and Technology, Daejeon 34141, Republic of Korea}

\author{Yong-Hoon Cho}
\email{yhc@kaist.ac.kr}
\affiliation{Department of Physics, Korea Advanced Institute of Science and Technology (KAIST), Daejeon 34141, Republic of Korea}

\author{Igor S. Aranson}
\email{isa12@psu.edu}
\affiliation{Departments of Biomedical Engineering, Chemistry, and Mathematics, The Pennsylvania State University, University Park 16802, USA}

\author{Hyoungsoon Choi}
\email{h.choi@kaist.ac.kr}
\affiliation{Department of Physics, Korea Advanced Institute of Science and Technology (KAIST), Daejeon 34141, Republic of Korea}

\author{Anton V. Nalitov}
\affiliation{Abrikosov Center for Theoretical Physics, Moscow Institute of Physics and Technology, Dolgoprudnyi, Moscow Region 141701, Russia}
\affiliation{Russian Quantum Center, Skolkovo, Moscow, 121205, Russia}

\author{Sergei V. Koniakhin}
\affiliation{Russian Quantum Center, Skolkovo, Moscow, 121205, Russia}
\affiliation{Center for Theoretical Physics of Complex Systems, Institute for Basic Science (IBS), Daejeon 34126, Republic of Korea}

\date{\today}

\begin{abstract}

Dynamics of exciton-polariton condensates under continuous-wave incoherent Gaussian optical pumping is considered. 
It is shown that the conventional supercritical Stuart-Landau picture is invalid in a certain domain of the parameter space.
For strong polariton repulsion from the reservoir and relatively small pump spots, the dynamics is adequately described by the quintic Stuart-Landau equation.
The corresponding subcritical pitchfork bifurcation leads to condensate formation, accompanied by bistability between the trivial and nontrivial states over a finite pump-power range and a one-bit memory.
Further increase of the repulsion parameter or decrease of the spot size breaks down the perturbative approach and leads to a peculiar self-trapping regime with complex dynamics.
Experimental evidence of the emergence of the proposed behavior is provided.
Our findings can be used to design polaritonic setups that exploit the predicted memory effect.

\end{abstract}

\maketitle

\sect{Introduction}Exciton-polaritons are hybrid bosonic quasiparticles in semiconducting microstructures~\cite{kavokin2007microcavities}.
They are formed by strongly coupled electron-hole pairs and cavity photons and exhibit peculiar properties~\cite{hopfield1958,weisbuch1992}.
In particular, they inherit low effective masses from cavity photons and giant interaction-induced optical nonlinearities from the excitonic part.
Remarkably, due to their low effective mass, polaritons can form bosonic condensates at temperatures of the order of tens of kelvin~\cite{deng2002condensation,kasprzak2006bose,balili2007bose} or even at room temperature for certain materials~\cite{christopoulos2007room,su2017room,georgakilas2025room} and acquire superfluid properties ~\cite{carusotto2013quantum}. 
Being inherently open and dissipative, polaritonic systems are fruitful for exploring fundamental phenomena of non-Hermitian interacting systems, for instance, topological~\cite{nalitov2015polariton} and supersolid phases~\cite{trypogeorgos2025emerging}, quantum turbulence~\cite{gavrilov2016towards,koniakhin20202d}, physics of solitons ~\cite{amo2011polariton} and vortices~\cite{lagoudakis2008quantized,choi2022observation}.
The interest in the field of polaritonics is also stimulated by several prospective applications as polariton simulators~\cite{berloff2017realizing,lagoudakis2017polariton,alyatkin2024antiferromagnetic} and controllable energy transferring~\cite{ballarini2019polaritonics,georgiou2021ultralong,cargioli2024,pajunp2024}.

Multistability of polariton condensates was shown in the resonant excitation scheme for the condensate density \cite{gippius2004nonlinear} and polarization \cite{Gippius2007,Sarkar2010} due to spin anisotropy of interaction nonlinearities.
The latter plays the key role for spin bifurcations and bistability of incoherently pumped polariton condensates \cite{Li2015b,Dreismann2016}.
Similarly, angular momentum bistability can develop in optically trapped rotating condensates \cite{Ma2020}.
However, switchable bistability between ``on'' and ``off'' states of an incoherently pumped polariton laser was limited to the electric pumping configuration \cite{Amthor2015}.

In this Letter, we demonstrate polariton condensate bistability and hysteresis under nonresonant optical pumping, arising from a subcritical pitchfork bifurcation beyond the validity range of the conventionally employed cubic Stuart-Landau model.
This effect is specifically pronounced for ballistic condensates generated by tightly focused pump spots and is due to the strong condensate-reservoir repulsion, which causes instability and breakdown of the cubic model.
Moreover, the condensate emerges with a certain finite density, thereby depleting the reservoir and reducing the repulsion energy with the latter. 
For yet stronger repulsions and smaller spots, our perturbative, weakly nonlinear approach fails, and the condensate is subject to self-trapping~\cite{dominici2015real}, when it forms a dense compact state in real space with complex dynamics (so-called hole-burning effect~\cite{estrecho2018single}).
Our findings are supported with experimental data on the condensate blueshift and intensity for varying pump spot sizes and can lead to further advances in the creation of polaritonic devices, where the memory due to the predicted bistability plays an important role.

\sect{Semi-analytical approach}The standard way to theoretically study the dynamics of exciton-polariton condensates is to employ the mean-field driven-dissipative Gross-Pitaevskii equation (ddGPE), accompanied by the rate equation for the reservoir~\cite{carusotto2013quantum}:
\be \label{res1}
  \partial_t n_\text{R} &=& P(\m{r},t) - \gamma_\text{R} n_\text{R} - R(n_\text{R}) |\psi|^2, \\ \label{ddGPE1}
  i \hbar \partial_t \psi &=& \bigl\{ -(\hbar^2/2m) (1 - i\beta) \nabla^2 + g_\text{C} |\psi|^2 \\ &&+ g_\text{R} n_\text{R} + i \hbar [R(n_\text{R}) - \gamma_\text{C}]/2 \bigr\} \psi. \nn
\ee
Here, $n_\text{R}$ is the density of the reservoir particles, $R(n_R)$ is their relaxation rate to the condensate with wave-function $\psi$, $P(\m{r},t)$ represents incoherent laser pumping, $\gamma_\text{R,C}$ are loss rates for the reservoir and the condensate, respectively. Parameter $\beta$ governs the momentum-dependent polariton damping ~\cite{solnyshkov2014hybrid}. Finally, $g_\text{R,C}$ are the effective constants of polariton repulsion from the reservoir particles and other polaritons, respectively. Below, we assume $R(n_\text{R}) = R_\text{SC} n_\text{R}$ and $\beta >0$ (in the opposite case, there is a short-wavelength instability of the condensate, so $\beta>0$ tends to stabilize smooth solutions~\cite{utesov2025universal}). 



\begin{figure}
    \centering
    \includegraphics[width=0.8\linewidth]{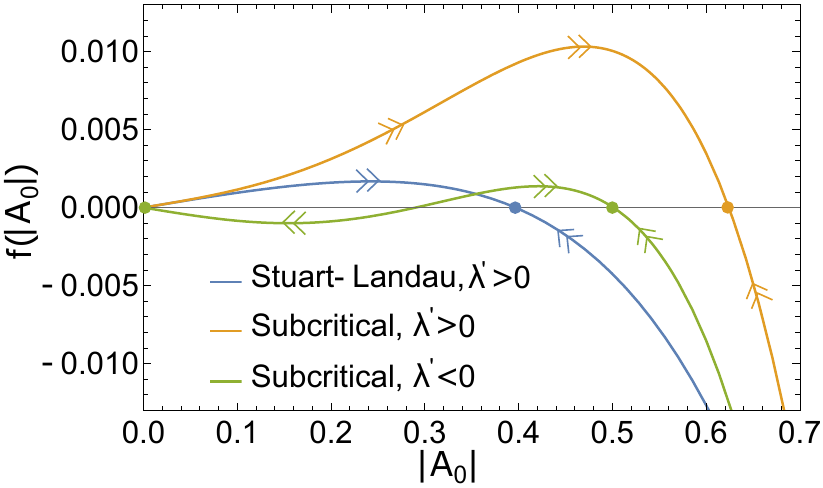}
    \caption{Sketch of the phase flow for the condensate amplitude $|A_0|$ according to Eqs.~\eqref{LS1} and~\eqref{LS3}. In the conventional Stuart-Landau picture for the single-spot exciton-polariton condensate above the pumping threshold, there is a single stable fixed point that determines its density. For the subcritical regime above the threshold, there is also a single stable fixed point. In contrast, slightly below the threshold, there are two stable fixed points: a non-trivial one with a nonzero condensate density and a trivial one with $|A_0|=0$, leading to bistability and memory effects. Upon further decrease of the pumping power, the nontrivial fixed point loses stability. }
    \label{FigFlow}
\end{figure}

\begin{figure*}[t]
  \centering
  \includegraphics[width=0.8\linewidth]{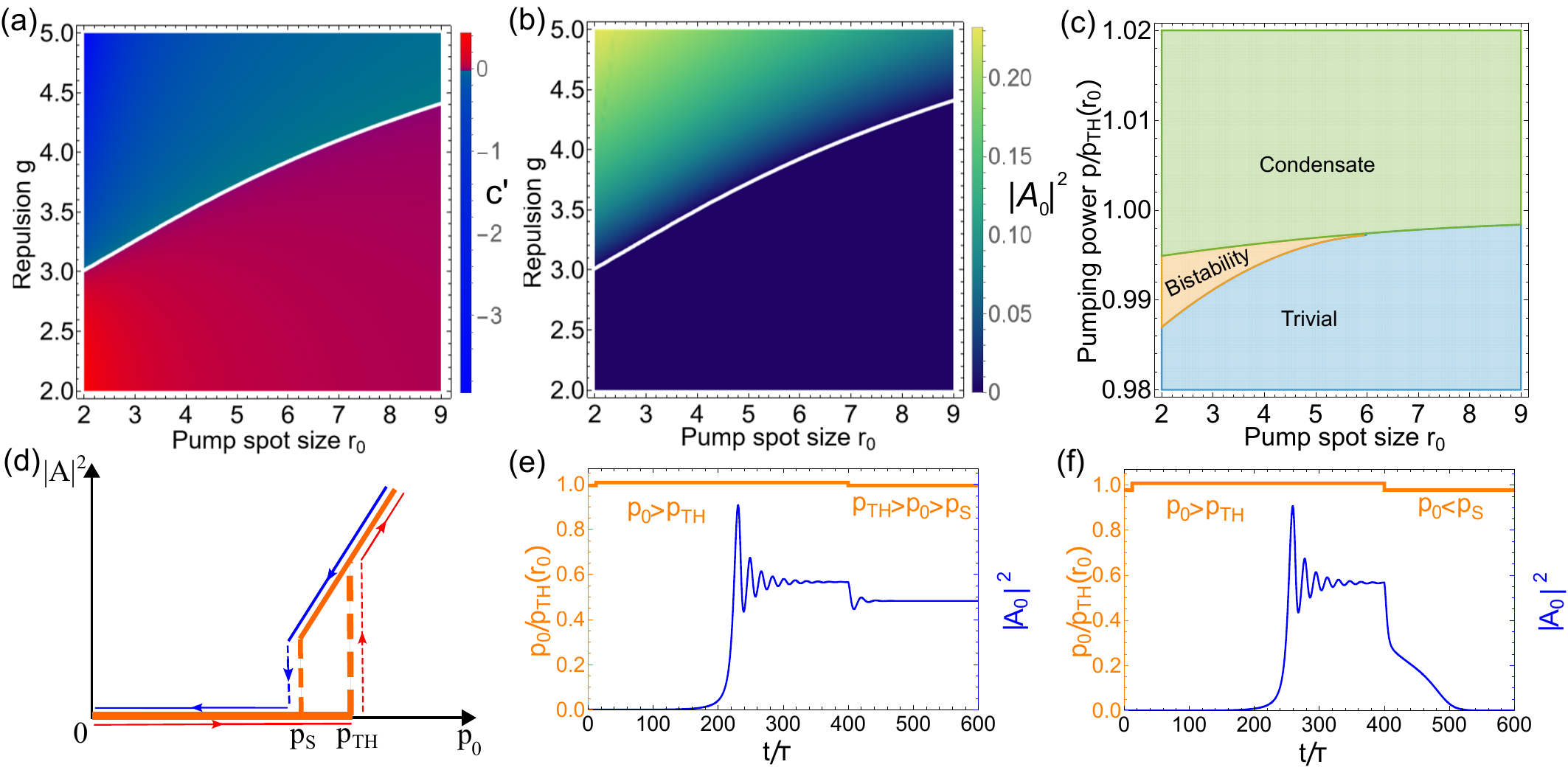}
  \caption{(a) Crucial parameter $\textrm{Re} \, c \equiv c^\prime$ [see Eqs.~\eqref{LS1} and~\eqref{LS3}] dependence on pumping spot size $r_0$ and repulsion parameter $g$, which measures the ratio between polariton-reservoir and polariton-polariton interaction strengths. The parameter plane can be divided into two parts with $c^\prime >0$, where the condensates can be discussed in terms of the Stuart-Landau equation, and with $c^\prime<0$, where there is a subcritical bifurcation and observable bistability. The white curve indicates the corresponding boundary.
  (b) Color map of the condensate density at the condensation threshold. For the blue region of panel (a), one observes $|A_0|^2 \neq 0$.
  (c) The condensation ``phase diagram.'' Here we fix $g=4$ and observe three regimes: trivial with $A_0 = 0$ shown in blue, nontrivial with $A_0 \neq 0$ shown in green, and bistability where either of the $A_0=0$ and $A_0 \neq 0$ states can exist shown in orange color.
  (d) Sketch of hysteresis for the polariton condensate for $c^\prime <0$. Subcritical bifurcation occurs at $p_\textrm{S}$, whereas the threshold pumping $p_\textrm{TH}$ here has the meaning of the required power to create condensate from the trivial $A=0$ state. Red lines indicate behavior upon an adiabatic increase in $p_0$, and blue lines indicate behavior upon a decrease.
  (e) and (f) Subcritical regime condensation for $r_0=3$ and $g=4$. Numerical $p_\textrm{TH}(r_0) \approx 8.99$. A state with random initial conditions for $A(\m{r})$ evolves into the condensate spot when pumping is tuned above $p_\textrm{TH}(r_0)$. After that, when making $p_0 < p_\textrm{TH}(r_0)$, the fate of the condensate drastically depends on $p_0$: in panel (e) $p_0 = 8.95$ and the condensate persists, whereas in panel (f) $p_0 = 8.8$ and the condensate density quickly goes to zero.
  In all panels other parameters are $\varkappa=1, \, \gamma=1,\,  \beta=0.1$. }    
  \label{FigSubcrit}
\end{figure*}

The system of equations~\eqref{res1} and~\eqref{ddGPE1} with certain parameters and pumping is suitable for direct numerical integration. However, to proceed (semi)analytically, further simplifications are in order. First, we use units related to the threshold for uniform pumping, $P_\textrm{TH} = \gamma_\text{C} \gamma_\text{R}/R_\textrm{SC}$~\cite{utesov2025universal}. The corresponding reservoir density is $n^{(0)}_\text{R} = R_\text{SC}/\gamma_\text{R}$ and we define the scales of length $\xi = \hbar/(2m g_\text{C} n^{(0)}_\text{R})^{1/2}$ and time $\tau = \hbar/g_\text{C} n^{(0)}_\text{R}$ accordingly. Then, we use dimensionless condensate wave function $A = \psi/(n^{(0)}_\text{R})^{1/2}$. Second, under the assumption of continuous wave (CW) pumping and slowly varying condensate density, we integrate out the reservoir
\be \label{res2}
  n_\text{R} = P(\m{r})/ \left(\gamma_\text{R} + R_\textrm{SC} |\psi|^2 \right),
\ee
and arrive at the equation
\be \label{ddGPE3}
  \partial_t A &=& \bigl\{ (i + \beta) \nabla^2   -i |A|^2 - i g P(\m{r})/ \left(1 + \gamma |A|^2\right) \nn \\ && + \varkappa \left[ P(\m{r})/ (1 + \gamma |A|^2) - 1 \right]/2 \bigr\} A, 
\ee
where the dimensionless parameters are
\be
  \varkappa = \hbar \gamma_\text{C}/g_\text{C} n^{(0)}_\text{R}, \quad \gamma = \gamma_\text{C}/\gamma_\text{R}, \quad g = g_\text{R}/g_\text{C}.
\ee

Following Ref.~\cite{utesov2025universal}, we can discuss physics near the pumping threshold (where $|A|^2 \ll 1$) using the expansion
\be \label{exp}
    P(\m{r})/(1 + \gamma |A|^2 ) \approx P(\m{r}) \left(1 - \gamma |A|^2\right),
\ee
which allows us to write the complex Ginzburg-Landau equation (cGLE)~\cite{aranson2002world}
\be \label{cGLE1}
  {\partial_t} A &=& \hat{\mathcal{L}} A - \left[ \left( \varkappa/2 - i g \right) \gamma P(\m{r})  + i \right] |A|^2 A,
\ee
with the linear operator
\be \label{lin1}
  \hat{\mathcal{L}} &=& (i + \beta) \nabla^2 + \left( \varkappa/2 - i g \right) P(\m{r}) - \varkappa/2.
\ee
For a Gaussian pumping $ P(r) = p_0 \exp{(-r^2/r^2_0)}$, we numerically seek for solutions of equation $\hat{\mathcal{L}} A = \lambda A$ with $A \propto e^{\lambda t + i M \varphi}$. Here $\lambda = \lambda^\prime - i \lambda^{\prime\prime}$ with $\lambda^\prime$ being the growth rate and $\lambda^{\prime\prime}$ being the condensate spectral line blueshift. Typically, the largest growth rate corresponds to a symmetrical mode with $M=0$; we denote its wave function as $\Psi_0$~\cite{utesov2025universal}. Using the weakly nonlinear analysis for $\lambda^\prime \ll 1$, we arrive at the Stuart-Landau equation for the condensate amplitude 
[$A(\m{r})= A_0(t) \Psi_0(r)$] 
\be \label{LS1}
  \partial_t A_0 = \lambda^\prime A_0 - c |A_0|^2 A_0,
\ee 
with 
\be \label{eqc}
  c &=& 2 \pi i \int^\infty_0 r |\Psi_0(r)|^2 \Psi^2_0(r)  dr + 2 \pi \gamma  \left( \frac{\varkappa}{2} - i g \right) \nn \\ &&\times \int^\infty_0  r P(r)|\Psi_0(r)|^2 \Psi^2_0(r) \equiv c^\prime + i c^{\prime \prime}.
\ee
Here normalization $ 2 \pi \int r  \Psi^2_0(r) dr = 1$ is assumed. The nontrivial fixed point $|A_0|=\sqrt{\lambda^\prime/c^\prime}$ emerges through a supercritical pitchfork bifurcation at the pumping threshold $p_{\textsc{th}}(r_0)$ and exists for $\lambda^\prime>0$ and $c^\prime>0$. However, the parameter $c$ is, in general, complex, and its real part $c^\prime$ need not be positive. 
Some intuition on $c^\prime$ can be obtained from the pseudo-pumping approximation $P(r) \approx p_0(1-r^2/r^2_0)$~\cite{utesov2025universal}, for which it was shown that it can be negative for $\beta>0$, small pumping spots, and large $g$, i.e., strong repulsion from the reservoir (see also End Matter for details). Results for the Gaussian pumping are shown in Fig.~\ref{FigSubcrit}(a), where $c^\prime<0$ in a large part of the plot.

The Stuart-Landau equation~\eqref{LS1} is inapplicable when $c^\prime <0 $ because it leads to an unbounded solution with $|A_0| \to \infty$ for $\lambda^\prime >0$. Note, however, that full cGLE can have bounded solutions even for $c^\prime<0$~\cite{schopf1991small}. In our case, we can try to cure the problem by including the next term in the expansion~\eqref{exp}.  This leads to the quintic Stuart-Landau equation (see End Matter)
\be \label{LS2}
  \partial_t A_0 = \lambda^\prime A_0 - c |A_0|^2 A_0 - d |A_0|^4 A_0,
\ee
where 
\be \label{eqd}
  d = - \gamma^2 \left( \frac{\varkappa}{2} - i g \right)  2 \pi \int^\infty_0   r P(r)|\Psi_0(r)|^4 \Psi^2_0(r)  dr.
\ee
Note that the term with the repulsion constant $g$ here has the opposite sign from its counterpart in Eq.~\eqref{eqc}, so we can expect that when $c^\prime<0$, we have $d^\prime >0$ and the solution for $A_0$ is bounded. Using particular parameters, we illustrate this idea in Figs.~\ref{FigSubcrit}(a) and (b). In the subsequent discussion, we will continue to call the pumping power for which $\lambda^\prime=0$ the ``condensation threshold'' despite the actual physics being more subtle.

\sect{Analysis of fixed points}For $d^\prime >0$, we proceed with Eq.~\eqref{LS2} analysis. Using the amplitude-phase representation $A_0 = |A_0| \exp{ (i \phi)}$, we arrive at the real equation:
\be
 \label{LS3}
  \partial_t |A_0| = \lambda^\prime |A_0| - c^\prime |A_0|^3 - d^\prime |A_0|^5 \equiv f(|A_0|).
\ee
The fixed points $f(|A_0|)=0$ analysis is trivial (see Fig.~\ref{FigFlow}): above the condensation threshold ($\lambda^\prime >0 $) there is only one stable fixed point 
\be \label{FP1}
  |A_0|^2 = \frac{\sqrt{|c^\prime|^2 + 4 d^\prime \lambda^\prime } + |c^\prime|}{2 d^\prime},
\ee
whereas below the threshold for $\lambda^\prime < 0$ and \mbox{$|c^\prime|^2 - 4 d^\prime |\lambda^\prime|>0$} there are two possibilities
\be \label{FP2}
  |A_0|^2 =0; \frac{\sqrt{|c^\prime|^2 - 4 d^\prime |\lambda^\prime| } + |c^\prime|}{2 d^\prime},
\ee
and for $\lambda^\prime < 0$ and \mbox{$|c^\prime|^2 - 4 d^\prime |\lambda^\prime| < 0$} only the trivial solution $A_0=0$ exists. We conclude that in the range of pumping powers corresponding to $\lambda^\prime \in [- |c^\prime|^2/ 4 d^\prime, 0]$, the phenomenon of \textit{bistability} emerges. This parameter subspace supports both the trivial state without a condensate and the nontrivial state with a finite condensate density. Importantly, the developed approach based on the quintic Stuart-Landau equation is formally valid for $|c^\prime|\ll 1$ only (see End Matter), so we expect the bistability region to be observable for quite a narrow pumping power range, as shown in Fig.~\ref{FigSubcrit}(c). There we fix $g=4$ and vary the spot size and the pumping power relative to the analytically predicted threshold~\eqref{APth}. According to Fig.~\ref{FigSubcrit}(a), we cross the line $c^\prime=0$ for $r_0 \approx 6$, and the bistability region emerges.

We see that the observable state of the polaritonic system essentially depends on the pumping history, and there is a hysteresis shown in Fig.~\ref{FigSubcrit}(d). Note that the density of the condensate at the pumping threshold is $|A_0|^2=|c^\prime|/d^\prime$, and at the subcritical bifurcation point, the density is half as large.

Using the approach developed above, it is not difficult to observe the bistability numerically. For a certain value of $g$, we chose the spot size below the one corresponding to the $c^\prime(r_0,g)=0$ curve [see Fig.~\ref{FigSubcrit}(a)]. Next, we used initial pumping with $p_0 < p_\textrm{TH}$; then, for some time, we increased it above the threshold and observed persistent condensate when returning to the initial $p_0$. If $p_0$ is taken outside the bistability range [see Fig.~\ref{FigSubcrit}(c)], the condensate asymptotically vanishes. Results of particular simulations are shown in Figs.~\ref{FigSubcrit}(e) and~(f).

\begin{figure}[t]
    \centering
    \includegraphics[width=0.8\linewidth]{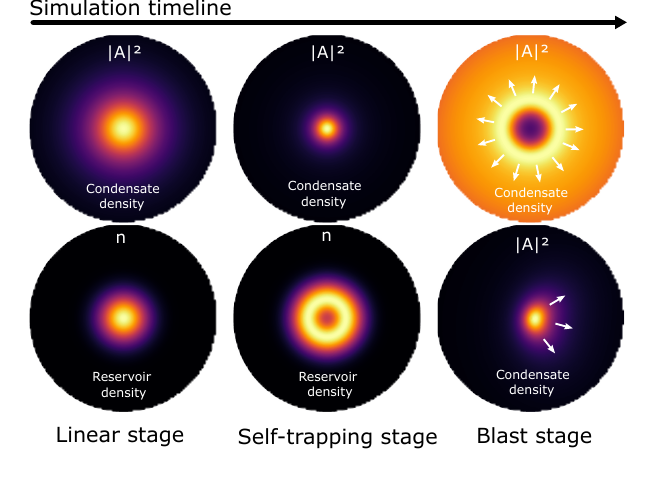}
    \caption{For yet smaller pumping spots and stronger repulsions from the reservoir, the ballistic condensate solution is subject to self-trapping. At the linear stage $|A|\ll1$, it grows following the pumping profile. When the condensate density is large enough, $\gamma |A|^2 \sim 1$, it ``collapses'' to a narrow spot, thus reducing the potential energy due to repulsion from the reservoir. The reservoir is essentially depleted in the center (so-called ``hole-burning'' effect). The self-trapped state is dynamically unstable. Depending on the model parameters, it can blast as a circular wave (upper-right panel) or ``tunnel'' from the trap (bottom-right panel) in a random direction.}
    \label{FigAuto}
\end{figure}

\sect{Condensate self-trapping}The subcritical regime discussed above relies on $\beta >0$ and $|c^\prime| \ll 1$ conditions. Fixing $\beta$, we can move further away from the $c^\prime =0$ line. The predicted value of $|A_0|^2$ increases. When it reaches a value $\sim 1/\gamma$, our perturbative approach is insufficient [cf. Eq.~\eqref{exp}]. The condensate density is high enough, and reservoir-mediated self-attraction becomes so strong that the condensate spatially separates from the reservoir, turning into a compact, dense spot. 
Below we briefly discuss our numerical observations; see Fig.~\ref{FigAuto} and~\cite{SM} for details.

Above the pumping threshold, i.e., $\lambda^\prime >0$, starting from a random initial condition, the condensate begins to grow, following the Gaussian pumping shape (linear stage). Then, when $|A_0|^2 \sim 1/\gamma$ is reached, it collapses and forms a narrow dense spot at the pumping spot center (self-trapping stage). The reservoir is essentially depleted there (``hole burning'' effect~\cite{estrecho2018single}).


Next, there is what we call the ``blast stage''. We observed that the fate of the self-trapped state depends on the pumping spot. For smaller spots, the condensate quickly ``explodes'' into a circular shock wave that carries the polaritons away (supernova-blast scenario). For larger spots and stronger pumps, the localized condensate can be observed for a long time as a metastable state. 
Eventually, it ``tunnels'' away from the trap in a certain direction and decays outside the pumping region. 

Finally, for $\lambda^\prime > 0$, the system starts again with the linear stage. One can also exploit the memory effect: for $\lambda^\prime < 0 $,  to observe the effects discussed above in the bistability range of pumpings, one needs to start from the state with sufficient condensate density. In this case, after the self-trapping and the blast stages, the condensate does not reappear unless pumping is increased.

 
To conclude this section, we would like to point out that exactly the $\beta$ parameter, which suppresses short-wavelength instabilities, allows us to observe a controllable subcritical condensation regime discussed in the previous sections. It ``interpolates'' between usual Stuart-Landau and strongly nonlinear self-trapping regimes.


\begin{figure}
    \centering
    \includegraphics[width=0.9\linewidth]{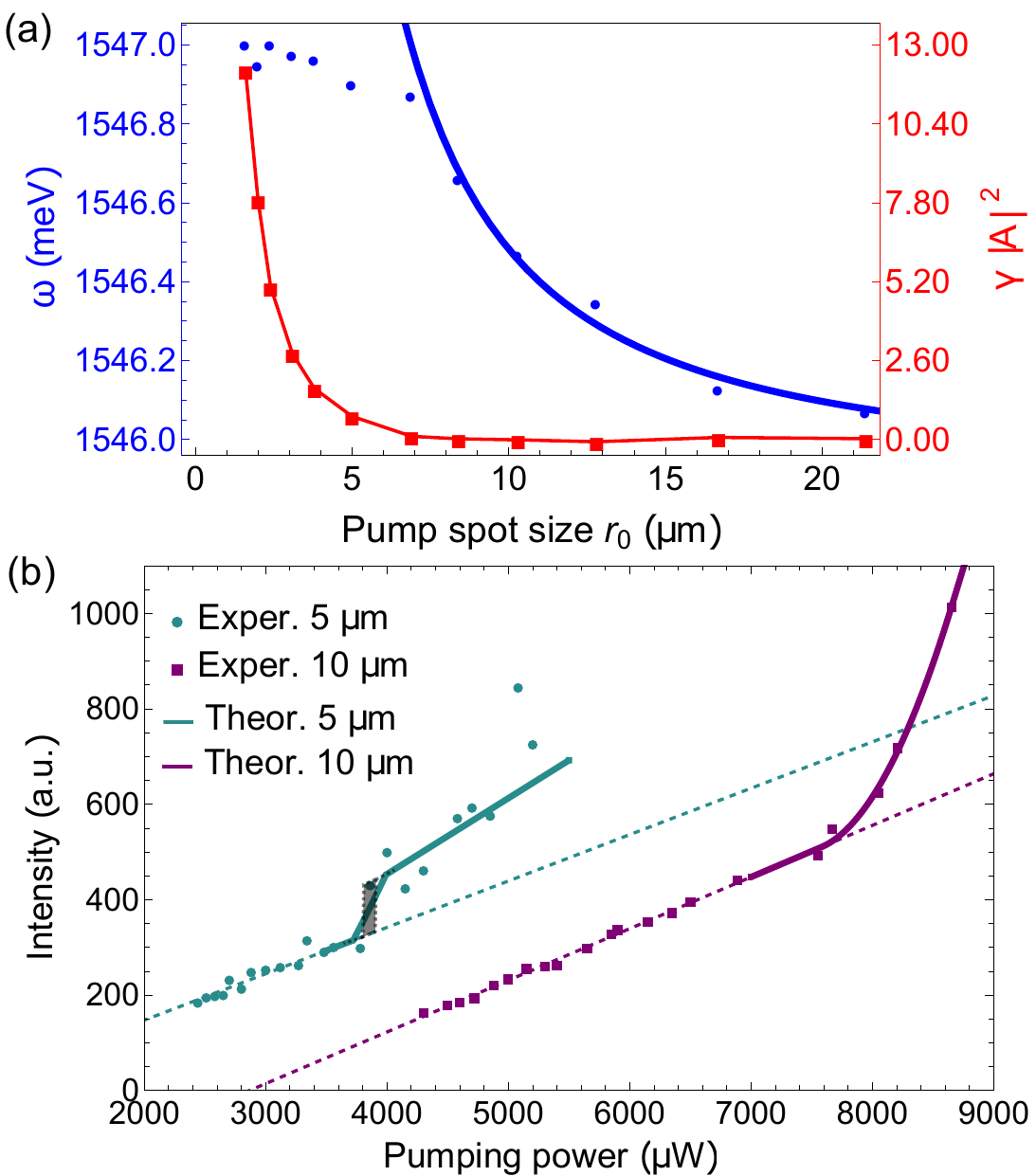}
    \caption{(a) Experimentally measured frequency of the condensate spectral line near the threshold as a function of the pumping spot size (blue dots) in an AlGaAs-based sample with $g \approx 4$ and the corresponding parameter of nonlinearity at the threshold $\gamma |A|^2$ (red dots) defined in Eq.~\eqref{BS}. Blue line indicates the frequency expected for the standard ``Stuart-Landau'' supercritical condensation. A crossover between large (quasi-homogeneous) and small (ballistic) spot cases is apparent. 
    (b) Spectral line intensity as a function of the pumping power (dots) and its interpretation in the model with fluctuating laser power (see End Matter). Dashed lines represent linear parts of the intensities beyond the condensation, providing a background signal. Gray dotted lines and shaded area illustrate hysteresis in a nearly ideal setup, cf. Fig.~\ref{FigSubcrit}(d). 
    }
    \label{FigExp}
\end{figure}

\sect{Experimental evidence} 
Direct observation of hysteresis and on-off switching for optical memory requires precise pump power control and suppression of fluctuations.
However, the presence of bistability near the subcritical pitchfork bifurcation can be inferred from standard emission measurements.
Specifically, we observe the transition to subcritical condensation for small pump spots in a pronounced deviation of the condensate energy from the universal linear scaling with pump spot size \cite{utesov2025universal}.
We find further evidence of the subcritical regime in the discontinuity of the emission intensity dependence on the pump power, which is typical for first-order phase transitions.

We used a Metal Organic Chemical Vapor Deposition GaAs-Al$_{0.15}$Ga$_{0.85}$As multiple quantum-well sample called KAIST 2 in our previous study~\cite{utesov2025universal} (see also~\cite{SM}) in the CW regime. Remarkably, an estimation from the Hopfield coefficients yields $g \approx 4$ for this ``photonic'' negatively detuned sample, making it suitable for testing subtle theoretical predictions. 

First, we examined the dependence of the condensate spectral line blueshift at the pumping threshold on the pumping spot size. The pumping threshold curve $p_\textrm{TH}(r_0)$ was accurately determined in our previous study~\cite{utesov2025universal}. It should also apply to more complex dynamical regimes such as the subcritical one, since condensation starts from the trivial $A_0=0$ state. For the standard Stuart-Landau (supercritical) regime, the main contribution to the blueshift is proportional to reservoir density at the spot center, $\Delta \omega \propto g_\textrm{R} n_\textrm{R}(\m{r}=0) \propto g p_\textrm{TH}(r_0)$ (see End Matter). In Fig.~\ref{FigExp}(a), we show that this relation holds only for large enough spots; for smaller ones, there is a significant deviation. Notably, these sizes belong to different condensation regimes: ballistic and quasi-homogeneous, since the characteristic length scale obtained previously is $R_0 \approx 7$~$\mu$m~\cite{utesov2025universal}. Phenomenologically, we can modify the blueshift formula assuming finite condensate density at the threshold (as in the subcritical regime):
\be \label{BS}
  \Delta \omega \propto g p_\textrm{TH}(r_0)/(1+ \gamma |A_0|^2).
\ee
The corresponding denominator correction $\gamma |A_0|^2$ due to the reservoir depletion [cf. Eq.~\eqref{res2}] varies from being negligible to values $\sim 10$ upon the spot size decrease. This parallels our theoretical expectations and hints at the subcritical regime for measured intermediate spot sizes $5$ and $6.9~\mu$m, where $\gamma |A_0|^2 \approx 0.10$ and $0.76$, respectively.

Second, in Fig.~\ref{FigExp}(b), we compare measurements of the spectral line intensity for $5$ and $10~\mu$m spots, which we expect to belong to different condensation regimes. We interpret their apparently different shape in terms of the model, accounting for the laser power time variation and the hysteresis loop, see End Matter and Supplementary Information~\cite{SM}. Notably, it predicts qualitatively different behavior in supercritical and subcritical cases. In the former, the linear growth of the condensate spectral line intensity is slightly smeared out near the threshold, and the curvature is positive. In contrast, in the subcritical regime, one can imagine a smearing of the hysteresis shown in Fig.~\ref{FigSubcrit}(d) where the crucial role is played by $|A_0|^2$ at the threshold. As the result, the curvature is negative.

\sect{Conclusions}
We show that on-off bistability emerges in nonequilibrium polariton condensates incoherently pumped with tightly focused optical beams, for which the conventional weakly nonlinear Stuart-Landau picture fails.
In this strongly nonlinear regime, reservoir depletion suppresses otherwise dominant ballistic losses, causing a positive response loop and instability of the cubic model.
Description of the stable condensate thus requires supplementing the Stuart-Landau with quintic corrections, which result in a subcritical pitchfork bifurcation and the aforementioned on-off bistability.
Further increase of the reservoir repulsion parameter or decrease in the pumping spot size leads to peculiar unstable and metastable self-trapped condensate states.
The direct observation of the memory effect requires high-precision control of pumping laser power, and observation of higher-power dynamical regimes is even more complicated due to their non-repeatability. However, the subcritical bifurcation and optical hysteresis fully explain the difference of blueshift and intensity power dependencies for small and large condensates observed in the GaAs/AlGaAs sample.
The proposed bistability and memory effect can become a physical ground for diverse polaritonic devices and setups, such as graph simulators, which exploit incoherent pumping.

\sect{Acknowledgements} We are grateful to Boris Altshuler for valuable discussions. 
O.I.U. and S.K.K. were supported by Brain Pool Plus Program through the National Research Foundation of Korea funded by the Ministry of Science and ICT (2020H1D3A2A03099291), National Research Foundation of Korea (NRF) grant funded by the Korea government (MSIT) (RS-2026-25470048), and Creation of the quantum information science R\&D ecosystem (based on human resources) through the National Research Foundation of Korea (NRF) funded by the Korean government (Ministry of Science and ICT (MSIT)) (RS-2023-00256050).
S.C. and H.C. acknowledge support from NRF (RS-2022-NR070358).
P.K. acknowledges support from St. Petersburg State University (Grant No. 125022803069-4).
A.N. and P.K. acknowledge support by the Russian Science Foundation under Grant No. 25-12-00135.


\bibliography{bib}

@article{balili2007bose,
  title={Bose-Einstein condensation of microcavity polaritons in a trap},
  author={Balili, Ryan and Hartwell, V and Snoke, David and Pfeiffer, L and West, Kayte},
  journal={Science},
  volume={316},
  number={5827},
  pages={1007--1010},
  year={2007},
  publisher={American Association for the Advancement of Science}
}

@article{choi2022realization,
  title={Realization of exciton-polariton condensation in GaAs-based microcavity grown by metalorganic chemical vapor deposition},
  author={Choi, Daegwang and Park, Min and Sung, Chan-Young and Choi, Hyoungsoon and Cho, Yong-Hoon},
  journal={Physical Review Research},
  volume={4},
  number={3},
  pages={033188},
  year={2022},
  publisher={APS}
}

@article{estrecho2018single,
  title={Single-shot condensation of exciton polaritons and the hole burning effect},
  author={Estrecho, E and Gao, Tingge and Bobrovska, Nataliya and Fraser, Michael D and Steger, M and Pfeiffer, L and West, K and Liew, TCH and Matuszewski, Michal and Snoke, David W and others},
  journal={Nature communications},
  volume={9},
  number={1},
  pages={2944},
  year={2018},
  publisher={Nature Publishing Group UK London}
}

@article{dominici2015real,
  title={Real-space collapse of a polariton condensate},
  author={Dominici, L and Petrov, M and Matuszewski, M and Ballarini, D and De Giorgi, Milena and Colas, D and Cancellieri, E and Silva Fern{\'a}ndez, B and Bramati, A and Gigli, Giuseppe and others},
  journal={Nature Communications},
  volume={6},
  number={1},
  pages={8993},
  year={2015},
  publisher={Nature Publishing Group UK London}
}

@article{weisbuch1992,
  title = {Observation of the coupled exciton-photon mode splitting in a semiconductor quantum microcavity},
  author = {Weisbuch, C. and Nishioka, M. and Ishikawa, A. and Arakawa, Y.},
  journal = {Phys. Rev. Lett.},
  volume = {69},
  issue = {23},
  pages = {3314--3317},
  numpages = {0},
  year = {1992},
  month = {Dec},
  publisher = {American Physical Society},
  doi = {10.1103/PhysRevLett.69.3314},
  url = {https://link.aps.org/doi/10.1103/PhysRevLett.69.3314}
}

@article{deng2002condensation,
  title={Condensation of semiconductor microcavity exciton polaritons},
  author={Deng, Hui and Weihs, Gregor and Santori, Charles and Bloch, Jacqueline and Yamamoto, Yoshihisa},
  journal={Science},
  volume={298},
  number={5591},
  pages={199--202},
  year={2002},
  publisher={American Association for the Advancement of Science}
}

@article{christopoulos2007room,
  title={Room-temperature polariton lasing in semiconductor microcavities},
  author={Christopoulos, S and Von H{\"o}gersthal, G Baldassarri H{\"o}ger and Grundy, AJD and Lagoudakis,  PG and Kavokin, AV and Baumberg, JJ and Christmann, G and Butt{\'e}, R and Feltin, E and Carlin, J-F and others},
  journal={Physical review letters},
  volume={98},
  number={12},
  pages={126405},
  year={2007},
  publisher={APS}
}

@article{ballarini2019polaritonics,
  title={Polaritonics: from microcavities to sub-wavelength confinement},
  author={Ballarini, Dario and De Liberato, Simone},
  journal={Nanophotonics},
  volume={8},
  number={4},
  pages={641--654},
  year={2019},
  publisher={De Gruyter}
}

@article{nalitov2015polariton,
  title={Polariton Z topological insulator},
  author={Nalitov, AV and Solnyshkov, DD and Malpuech, G},
  journal={Physical review letters},
  volume={114},
  number={11},
  pages={116401},
  year={2015},
  publisher={APS}
}

@article{solnyshkov2014hybrid,
  title={Hybrid {B}oltzmann--{G}ross-{P}itaevskii theory of {B}ose-{E}instein condensation and superfluidity in open driven-dissipative systems},
  author={Solnyshkov, DD and Tercas, H and Dini, K and Malpuech, G},
  journal={Physical Review A},
  volume={89},
  number={3},
  pages={033626},
  year={2014},
  publisher={APS}
}

@article{aranson2002world,
  title={The world of the complex Ginzburg-Landau equation},
  author={Aranson, Igor S and Kramer, Lorenz},
  journal={Reviews of modern physics},
  volume={74},
  number={1},
  pages={99},
  year={2002},
  publisher={APS}
}

@article{lagoudakis2008quantized,
  title={Quantized vortices in an exciton--polariton condensate},
  author={Lagoudakis, Konstantinos G and Wouters, Michiel and Richard, Maxime and Baas, Augustin and Carusotto, Iacopo and Andr{\'e}, Regis and Dang, Le Si and Deveaud-Pl{\'e}dran, B},
  journal={Nature physics},
  volume={4},
  number={9},
  pages={706--710},
  year={2008},
  publisher={Nature Publishing Group UK London}
}

@article{choi2022observation,
  title={Observation of a single quantized vortex vanishment in exciton-polariton superfluids},
  author={Choi, Daegwang and Park, Min and Oh, Byoung Yong and Kwon, Min-Sik and Park, Suk In and Kang, Sooseok and Song, Jin Dong and Ko, Dogyun and Sun, Meng and Savenko, Ivan G and others},
  journal={Physical Review B},
  volume={105},
  number={6},
  pages={L060502},
  year={2022},
  publisher={APS}
}

@article{carusotto2013quantum,
  title={Quantum fluids of light},
  author={Carusotto, Iacopo and Ciuti, Cristiano},
  journal={Reviews of Modern Physics},
  volume={85},
  number={1},
  pages={299--366},
  year={2013},
  publisher={APS}
}

@article{Smirnov2014dynamics,
  title = {Dynamics and stability of dark solitons in exciton-polariton condensates},
  author = {Smirnov, Lev A. and Smirnova, Daria A. and Ostrovskaya, Elena A. and Kivshar, Yuri S.},
  journal = {Phys. Rev. B},
  volume = {89},
  issue = {23},
  pages = {235310},
  numpages = {11},
  year = {2014},
  month = {Jun},
  publisher = {American Physical Society},
  doi = {10.1103/PhysRevB.89.235310},
  url = {https://link.aps.org/doi/10.1103/PhysRevB.89.235310}
}

@article{hopfield1958,
  title = {Theory of the Contribution of Excitons to the Complex Dielectric Constant of Crystals},
  author = {Hopfield, J. J.},
  journal = {Phys. Rev.},
  volume = {112},
  issue = {5},
  pages = {1555--1567},
  numpages = {0},
  year = {1958},
  month = {Dec},
  publisher = {American Physical Society},
  doi = {10.1103/PhysRev.112.1555},
  url = {https://link.aps.org/doi/10.1103/PhysRev.112.1555}
}

@article{kavokin2007microcavities,
  title={Microcavities Oxford University Press Inc},
  author={Kavokin, A and Baumberg, JJ and Malpuech, G and Laussy, FP},
  journal={New York},
  year={2007}
}

@article{kasprzak2006bose,
  title={Bose--Einstein condensation of exciton polaritons},
  author={Kasprzak, Jacek and Richard, Murielle and Kundermann, S and Baas, A and Jeambrun, P and Keeling, Jonathan Mark James and Marchetti, FM and Szyma{\'n}ska, MH and Andr{\'e}, R and Staehli, JL and others},
  journal={Nature},
  volume={443},
  number={7110},
  pages={409--414},
  year={2006},
  publisher={Nature Publishing Group UK London}
}

@article{georgiou2021ultralong,
  title={Ultralong-range polariton-assisted energy transfer in organic microcavities},
  author={Georgiou, Kyriacos and Jayaprakash, Rahul and Othonos, Andreas and Lidzey, David G},
  journal={Angewandte Chemie},
  volume={133},
  number={30},
  pages={16797--16803},
  year={2021},
  publisher={Wiley Online Library}
}

@article{cargioli2024,
url = {https://doi.org/10.1515/nanoph-2023-0677},
title = {Active control of polariton-enabled long-range energy transfer},
author = {Alessio Cargioli and Maksim Lednev and Lorenzo Lavista and Andrea Camposeo and Adele Sassella and Dario Pisignano and Alessandro Tredicucci and Francisco J. Garcia-Vidal and Johannes Feist and Luana Persano},
pages = {2541--2551},
volume = {13},
number = {14},
journal = {Nanophotonics},
doi = {doi:10.1515/nanoph-2023-0677},
year = {2024},
lastchecked = {2025-01-13}
}

@article{pajunp2024,
  title = {Polariton-assisted long-distance energy transfer between excitons in two-dimensional semiconductors},
  author = {Pajunp\"a\"a, Tuomas and Nigmatulin, Fedor and Akkanen, Suvi-Tuuli and Fernandez, Henry and Groenhof, Gerrit and Sun, Zhipei},
  journal = {Phys. Rev. B},
  volume = {109},
  issue = {19},
  pages = {195409},
  numpages = {7},
  year = {2024},
  month = {May},
  publisher = {American Physical Society},
  doi = {10.1103/PhysRevB.109.195409},
  url = {https://link.aps.org/doi/10.1103/PhysRevB.109.195409}
}

@article{berloff2017realizing,
  title={Realizing the classical XY Hamiltonian in polariton simulators},
  author={Berloff, Natalia G and Silva, Matteo and Kalinin, Kirill and Askitopoulos, Alexis and T{\"o}pfer, Julian D and Cilibrizzi, Pasquale and Langbein, Wolfgang and Lagoudakis, Pavlos G},
  journal={Nature materials},
  volume={16},
  number={11},
  pages={1120--1126},
  year={2017},
  publisher={Nature Publishing Group UK London}
}

@article{lagoudakis2017polariton,
  title={A polariton graph simulator},
  author={Lagoudakis, Pavlos G and Berloff, Natalia G},
  journal={New Journal of Physics},
  volume={19},
  number={12},
  pages={125008},
  year={2017},
  publisher={IOP Publishing}
}

@article{alyatkin2024antiferromagnetic,
  title={Antiferromagnetic Ising model in a triangular vortex lattice of quantum fluids of light},
  author={Alyatkin, Sergey and Mili{\'a}n, Carles and Kartashov, Yaroslav V and Sitnik, Kirill A and Gnusov, Ivan and T{\"o}pfer, Julian D and Sigur{\dh}sson, Helgi and Lagoudakis, Pavlos G},
  journal={Science Advances},
  volume={10},
  number={34},
  pages={eadj1589},
  year={2024},
  publisher={American Association for the Advancement of Science}
}

@article{amo2011polariton,
  title={Polariton superfluids reveal quantum hydrodynamic solitons},
  author={Amo, Alberto and Pigeon, S and Sanvitto, D and Sala, VG and Hivet, R and Carusotto, Iacopo and Pisanello, F and Lem{\'e}nager, G and Houdr{\'e}, R and Giacobino, E and others},
  journal={Science},
  volume={332},
  number={6034},
  pages={1167--1170},
  year={2011},
  publisher={American Association for the Advancement of Science}
}

@article{koniakhin20202d,
  title={2D quantum turbulence in a polariton quantum fluid},
  author={Koniakhin, SV and Bleu, O and Malpuech, G and Solnyshkov, DD},
  journal={Chaos, Solitons \& Fractals},
  volume={132},
  pages={109574},
  year={2020},
  publisher={Elsevier}
}

@article{gavrilov2016towards,
  title={Towards spin turbulence of light: Spontaneous disorder and chaos in cavity-polariton systems},
  author={Gavrilov, SS},
  journal={Physical Review B},
  volume={94},
  number={19},
  pages={195310},
  year={2016},
  publisher={APS}
}

@article{utesov2025universal,
  title={Universal condensation threshold dependence on pump beam size for exciton-polaritons},
  author={Utesov, Oleg I and Park, Min and Choi, Daegwang and Choi, Soohong and Park, Suk In and Kang, Sooseok and Song, Jin Dong and Osipov, Alexey N and Yulin, Alexey V and Cho, Yong-Hoon and others},
  journal={Communications Physics},
  volume={8},
  number={1},
  pages={286},
  year={2025},
  publisher={Nature Publishing Group UK London}
}

@article{georgakilas2025room,
  title={Room-temperature cavity exciton-polariton condensation in perovskite quantum dots},
  author={Georgakilas, Ioannis and Tiede, David and Urbonas, Darius and Mirek, Rafa{\l} and Bujalance, Clara and Cali{\`o}, Laura and Oddi, Virginia and Tao, Rui and Dirin, Dmitry N and Rain{\`o}, Gabriele and others},
  journal={Nature Communications},
  volume={16},
  number={1},
  pages={5228},
  year={2025},
  publisher={Nature Publishing Group UK London}
}

@article{su2017room,
  title={Room-temperature polariton lasing in all-inorganic perovskite nanoplatelets},
  author={Su, Rui and Diederichs, Carole and Wang, Jun and Liew, Timothy CH and Zhao, Jiaxin and Liu, Sheng and Xu, Weigao and Chen, Zhanghai and Xiong, Qihua},
  journal={Nano letters},
  volume={17},
  number={6},
  pages={3982--3988},
  year={2017},
  publisher={ACS Publications}
}

@article{trypogeorgos2025emerging,
  title={Emerging supersolidity in photonic-crystal polariton condensates},
  author={Trypogeorgos, Dimitrios and Gianfrate, Antonio and Landini, Manuele and Nigro, Davide and Gerace, Dario and Carusotto, Iacopo and Riminucci, Fabrizio and Baldwin, Kirk W and Pfeiffer, Loren N and Martone, Giovanni I and others},
  journal={Nature},
  volume={639},
  number={8054},
  pages={337--341},
  year={2025},
  publisher={Nature Publishing Group UK London}
}

@article{gippius2004nonlinear,
  title={Nonlinear dynamics of polariton scattering in semiconductor microcavity: Bistability vs. stimulated scattering},
  author={Gippius, NA and Tikhodeev, SG and Kulakovskii, VD and Krizhanovskii, DN and Tartakovskii, AI},
  journal={EPL (Europhysics Letters)},
  volume={67},
  number={6},
  pages={997--1003},
  year={2004}
}

@Article{Sarkar2010,
  author    = {Sarkar, D. and Gavrilov, S. S. and Sich, M. and Quilter, J. H. and Bradley, R. A. and Gippius, N. A. and Guda, K. and Kulakovskii, V. D. and Skolnick, M. S. and Krizhanovskii, D. N.},
  journal   = {Physical Review Letters},
  title     = {Polarization Bistability and Resultant Spin Rings in Semiconductor Microcavities},
  year      = {2010},
  issn      = {1079-7114},
  month     = nov,
  number    = {21},
  pages     = {216402},
  volume    = {105},
  doi       = {10.1103/physrevlett.105.216402},
  publisher = {American Physical Society (APS)},
}

@Article{Gippius2007,
  author    = {Gippius, N. A. and Shelykh, I. A. and Solnyshkov, D. D. and Gavrilov, S. S. and Rubo, Yuri G. and Kavokin, A. V. and Tikhodeev, S. G. and Malpuech, G.},
  journal   = {Physical Review Letters},
  title     = {Polarization Multistability of Cavity Polaritons},
  year      = {2007},
  issn      = {1079-7114},
  month     = jun,
  number    = {23},
  pages     = {236401},
  volume    = {98},
  doi       = {10.1103/physrevlett.98.236401},
  publisher = {American Physical Society (APS)},
}

@Article{Li2015b,
  author    = {Li, G. and Liew, T. C. H. and Egorov, O. A. and Ostrovskaya, E. A.},
  journal   = {Physical Review B},
  title     = {Incoherent excitation and switching of spin states in exciton-polariton condensates},
  year      = {2015},
  issn      = {1550-235X},
  month     = aug,
  number    = {6},
  pages     = {064304},
  volume    = {92},
  doi       = {10.1103/physrevb.92.064304},
  publisher = {American Physical Society (APS)},
}

@Article{Dreismann2016,
  author    = {Dreismann, Alexander and Ohadi, Hamid and del Valle-Inclan Redondo, Yago and Balili, Ryan and Rubo, Yuri G. and Tsintzos, Simeon I. and Deligeorgis, George and Hatzopoulos, Zacharias and Savvidis, Pavlos G. and Baumberg, Jeremy J.},
  journal   = {Nature Materials},
  title     = {A sub-femtojoule electrical spin-switch based on optically trapped polariton condensates},
  year      = {2016},
  issn      = {1476-4660},
  month     = aug,
  number    = {10},
  pages     = {1074--1078},
  volume    = {15},
  doi       = {10.1038/nmat4722},
  publisher = {Springer Science and Business Media LLC},
}

@Article{Ma2020,
  author    = {Ma, Xuekai and Berger, Bernd and Aßmann, Marc and Driben, Rodislav and Meier, Torsten and Schneider, Christian and Höfling, Sven and Schumacher, Stefan},
  journal   = {Nature Communications},
  title     = {Realization of all-optical vortex switching in exciton-polariton condensates},
  year      = {2020},
  issn      = {2041-1723},
  month     = feb,
  number    = {1},
  volume    = {11},
  doi       = {10.1038/s41467-020-14702-5},
  publisher = {Springer Science and Business Media LLC},
}

@Article{Amthor2015,
  author    = {Amthor, M. and Liew, T. C. H. and Metzger, C. and Brodbeck, S. and Worschech, L. and Kamp, M. and Shelykh, I. A. and Kavokin, A. V. and Schneider, C. and Höfling, S.},
  journal   = {Physical Review B},
  title     = {Optical bistability in electrically driven polariton condensates},
  year      = {2015},
  issn      = {1550-235X},
  month     = feb,
  number    = {8},
  pages     = {081404},
  volume    = {91},
  doi       = {10.1103/physrevb.91.081404},
  publisher = {American Physical Society (APS)},
}

@misc{SM,
  note = {See Supplemental Material at [URL to be inserted by publisher] for details of numerical calculations  and experimental sample.}
}

@article{schopf1991small,
  title={Small-amplitude periodic and chaotic solutions of the complex Ginzburg-Landau equation for a subcritical bifurcation},
  author={Sch{\"o}pf, Wolfgang and Kramer, Lorenz},
  journal={Physical review letters},
  volume={66},
  number={18},
  pages={2316},
  year={1991},
  publisher={APS}
}

\begin{widetext}
    \newpage
\end{widetext}

\begin{center}

\LARGE \bf End Matter
    
\end{center}

\sect{Condensate under Gaussian pumping}Here we discuss some important properties of the condensates under the Gaussian pumping $ P(r) = p_0 \exp{(-r^2/r^2_0)}$ near the threshold, which stem from the solution presented in Ref.~\cite{utesov2025universal}. The eigenvalue of the condensing state $\Psi_0$ reads
\be \label{AlamH}
  \lambda = (p_0 - 1) \frac{\varkappa}{2} - i g p_0  - 2 \sqrt{(i+\beta)\left(\frac{\varkappa}{2} - i g \right) \frac{p_0}{r^2_0}}.
\ee
We express it through real and imaginary parts as $\lambda = \lambda^\prime - i \lambda^{\prime \prime}$. The pumping threshold corresponds to $\lambda^\prime=0$. For the spot of the size $r_0$ reads
\be \label{APth}
  p_{\textsc{th}}(r_0) = \left( R_0/ r_0 + \sqrt{R^2_0/ r^2_0+1}\right)^2,
\ee
where
\be
  R_0 = 2 \, \textrm{Re}\left[ \sqrt{(i+\beta) \left(\varkappa/2 - i g \right)} \right]/\varkappa
\ee
is the characteristic length scale for a given sample. It was shown that $r_0 \gg R_0$ corresponds to the quasi-homogeneous pumping regime, whereas for $r_0 \lesssim R_0$ so-called ballistic condensation occurs, where the main loss mechanism is due to runaway polaritons. In the former case, $p_0 \sim 1$ and for $g \gg 1$ we have
\be
  |\Psi_0(\m{r})|  \approx e^{-r^2/4 r_0 R_0},
\ee
which means that the effective size of the condensate droplet is $\sim \sqrt{r_0 R_0} \ll r_0$. So, even though for $g \gg 1$ we are in the focusing limit of the nonlinear equation~\eqref{cGLE1} for the homogeneous pumping~\cite{Smirnov2014dynamics}, our solution is already compact in space and is not subject to self-trapping for $p_0 \sim p_{\textsc{th}}$. It is also apparent from the estimation $c^\prime \approx \gamma \varkappa/2$ for $r_0 \gg R_0$, which indicates that the Stuart-Landau equation~\eqref{LS1} is applicable.

In another case, which is of particular interest to us, the estimations stem from relations
\be
  p_0 \approx 4 R^2_0/r^2_0 \gg 1, \quad R_0 \approx 2 \sqrt{g}/\varkappa.
\ee
They result in
\be
  |\Psi_0(\m{r})| \approx e^{-r^2/2 r^2_0},
\ee
which shows that the condensate density follows the pumping profile. Also, it is easy to see that the $g p_0$ term in $\lambda^{\prime \prime}$ mainly determines the condensate spectral line blueshift. Regarding the crucial parameter $c^\prime$, one can show that in the limit of $g \gg 1$ and other parameters $\sim 1$, the following estimation can be used:
\be
  c^\prime = - \frac{4 \beta \gamma}{\varkappa^2 r^2_0} g^2 + \frac{\gamma \varkappa}{2 r^2_0 g }+O(1/g^2) \approx - \frac{R^2_0}{r^2_0} \beta \gamma g.
\ee
Notably, the $O(1/g)$ term here is always positive, so to observe the subcritical regime one needs $\beta$ larger than a certain value.

\sect{Derivation of quintic Stuart-Landau equation}The conventional Stuart-Landau equation~\eqref{LS1} relies on the condition $\lambda^\prime \ll 1$ where nonlinearity is weak, $A^3_0 \sim (\lambda^\prime)^{3/2}$. In the subcritical case $\text{Re} \, c <0 $, the quintic Stuart-Landau equation~\eqref{LS2} can also be derived in the weakly nonlinear regime. Fixed point solutions~\eqref{FP1} and~\eqref{FP2} suggest taking advantage of the $|c^\prime| \ll 1$ condition. In particular, on general grounds, $|d| \sim 1$ and the relevant range of pumping powers corresponds to $|\lambda^\prime| \sim (c^\prime)^2$. 

The complex Ginzburg-Landau equation with higher-order terms reads [cf. Eq.~\eqref{cGLE1}]
\be \label{cGLE2}
  {\partial_t} A &=& \hat{\mathcal{L}} A - \left[ \left( \frac{\varkappa}{2} - i g \right) \gamma P(\m{r})  + i \right] |A|^2 A  \\ && +  \left( \frac{\varkappa}{2} - i g \right) \gamma^2 P(\m{r}) |A|^4 A \nn.
\ee
Following the general idea of weakly nonlinear analysis, we write
\be \label{AnlWF}
  A(\m{r},t) = \left[ A_0 (t) \Psi_{0}(\m{r}) + w(\m{r},t) \right] e^{- i \lambda^{\prime\prime} t},
\ee
and introduce $\hat{\mathcal{L}}^\prime = \hat{\mathcal{L}} + i \lambda^{\prime \prime}$. Now we can discuss the hierarchy of various quantities. The solution $A_0 \sim (c^\prime)^{1/2}$, all the terms in Eq.~\eqref{LS2} are $\sim (c^\prime)^{5/2}$, hence the correction generated by nonlinearity $w \sim (c^\prime)^{5/2}$, whereas its time derivative $\partial_t w \sim A^2_0 \partial_t A_0\sim (c^\prime)^{7/2}$ is negligible. Collecting the important terms in Eq.~\eqref{cGLE2}, we have
\be
  \hat{\mathcal{L}}^\prime w &=& \Psi_0 \partial_t A_0 - \lambda^\prime \Psi_0 A_0 \nn \\ &+& \left[ \left( \frac{\varkappa}{2} - i g \right) \gamma P(\m{r})  + i \right] |A_0|^2 A_0 |\Psi_0|^2 \Psi_0  \\  &-&   \left( \frac{\varkappa}{2} - i g \right) \gamma^2 P(\m{r}) |A_0|^4 A_0 |\Psi_0|^4 \Psi_0. \nn
\ee
Then, we multiply both sides by $\Psi_{0}$ and note that 
\be
  \int  \Psi_{0} \hat{\mathcal{L}}^\prime w d^2 \m{r} = \int w \hat{\mathcal{L}}^\prime \Psi_{0}  d^2 \m{r} \sim \lambda^\prime (c^\prime)^{5/2} 
\ee
is negligible (this is the solvability condition for our problem), which leads us to the quintic Stuart-Landau equation~\eqref{LS2} for condensate amplitude $A_0$ with certain coefficients. In simple terms, it is expected to work well near the pumping threshold and the curve $c^\prime = 0$ in the parameter space. In practice, our numerics indicate that both conditions can be relaxed to some extent, and Eq.~\eqref{LS2} still works semi-quantitatively.

We also note that the third-order nonlinearity modifies the shape of the condensate in real space due to admixture of other harmonics. Numerically, we observed that this effect is negligible in the weakly-nonlinear regime.

\sect{Phenomenological model with fluctuating laser}Laser power stabilization can be crucial for experimental observation of the theoretically predicted memory effect. Nevertheless, even if the pumping power fluctuates with time, a certain behavior of the condensate spectral line intensity can be predicted. We assume a box distribution of the instantaneous laser powers, $p^\prime \in (p-\Delta p, p + \Delta p)$, and that the corresponding time scale of $p^\prime$ variation is much longer than the condensate formation time $\lesssim 1$~ns. 

In the supercritical Stuart-Landau regime, we can use~\cite{utesov2025universal}  
\be
  I(p^\prime) \propto (p^\prime-p_\textsc{th}) \theta(p^\prime-p_\textsc{th}) 
\ee
and the observable intensity is given by
\be \label{Int1}
 \langle I(p) \rangle  &=& \frac{1}{2 \Delta p} \int^{p+\Delta p}_{p-\Delta p} I(p^\prime) dp^\prime \\ &=& \left\{
 \begin{array}{c}
    p-p_\textsc{th} ,  \quad p>p_\textsc{th} + \Delta p,  \\
    \frac{(p + \Delta p - p_\textsc{th})^2}{4\Delta p},  \quad  |p- p_\textsc{th}| < \Delta p, \\
    0,  \quad p < p_\textsc{th} - \Delta p.
  \end{array} 
\right. \nn 
\ee
Thus, the intensity growth is parabolic at $p_\textsc{th} - \Delta p$, and linear at $p>p_\textsc{th} + \Delta p$.

In the subcritical regime, for simplicity, we can discuss the case when $\Delta p > p_\textsc{th} - p_\textsc{s}$, and assign a certain intensity $I_0$ for $p^\prime = p_\textsc{th}$, so 
\be
  I(p^\prime) = I_0 + k (p^\prime-p_\textsc{th}) \theta(p^\prime-p_\textsc{th}),
\ee
where $k$ is some coefficient.
Equation~\eqref{Int1} is modified accordingly:
\be \label{Int2}
 \langle I(p) \rangle &=& \left\{
 \begin{array}{c}
    I_0 + k(p-p_\textsc{th}) ,  \quad p>p_\textsc{th} + \Delta p,  \\
    I_0\frac{(p + \Delta p - p_\textsc{th})}{2\Delta p}+ k \frac{(p + \Delta p - p_\textsc{th})^2}{4\Delta p}, \\  |p- p_\textsc{th}| < \Delta p, \\
    0,  \quad p < p_\textsc{th} - \Delta p.
  \end{array} 
\right. 
\ee
Notably, the resulting curve combines a quick increase of intensity near $p_\textsc{th}$, whereas the further growth for $p>p_\textsc{th} + \Delta p$ is slower.  This is a direct consequence of the hysteresis and finite condensate density at the threshold for the subcritical regime.


\clearpage
\renewcommand{\thefigure}{S\arabic{figure}}
\setcounter{figure}{0} 

\renewcommand\theequation{S\arabic{equation}}
\setcounter{equation}{0} 

\renewcommand{\thetable}{S\arabic{table}}
\setcounter{table}{0}

\renewcommand{\thesection}{\Roman{section}}
\renewcommand{\thesubsection}{\thesection.\arabic{subsection}}

\setcounter{section}{0}
\setcounter{subsection}{0}
\setcounter{secnumdepth}{2}

\renewcommand{\section}[1]{%
  \refstepcounter{section}%
  \setcounter{subsection}{0}%
  \par\bigskip
  \begin{center}
  {\bfseries Supplementary Note \thesection: #1}
  \end{center}
  \par\medskip
}

\renewcommand{\subsection}[1]{%
  \refstepcounter{subsection}%
  \par\medskip
  \begin{center}
  {\bfseries Section \thesubsection: #1}
  \end{center}
  \par\smallskip
}

\makeatletter
\renewcommand{\p@section}{}
\renewcommand{\p@subsection}{}
\makeatother

\onecolumngrid
\begin{center}
\Large{Supplementary Information}
\end{center}


\section{Numerical calculations}

In the weakly-nonlinear regime, we numerically obtained eigenmodes of the linear operator~\eqref{lin1} with Gaussian pumping to find a state with the highest growth rate $\Psi_0(r)$. Next, in line with our semi-analytical approach, we numerically calculated integrals for parameters $c$ and $d$ [Eqs.~\eqref{eqc} and~\eqref{eqd}].  

\begin{figure}[b]
    \centering
    \includegraphics[width=0.6\linewidth]{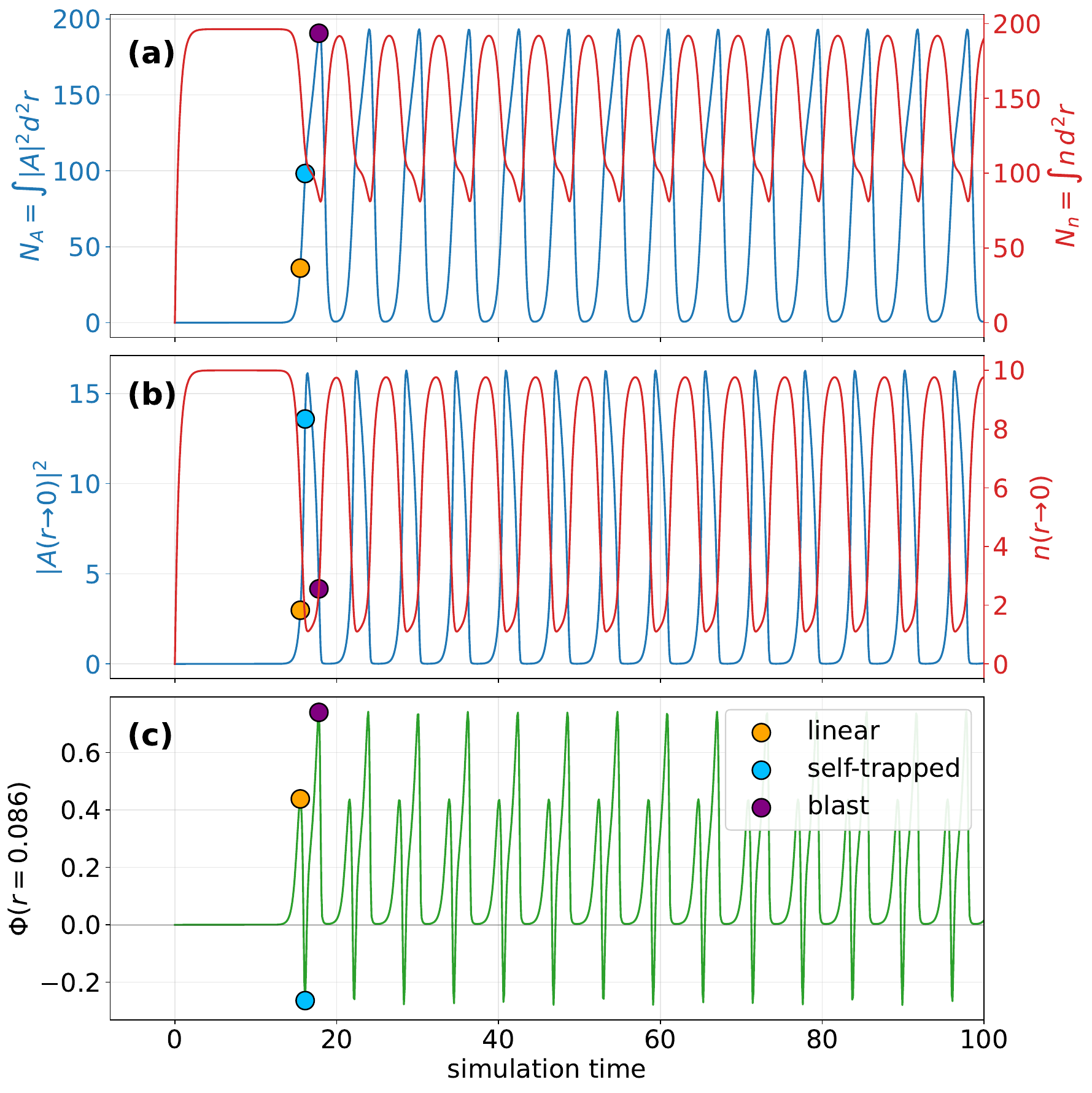}
    \caption{Supernova-blast scenario overview. Dynamics consists of a repeating cycle of condensate growth, self-trapping, and explosion; see also Fig.~3 of the Main Text. (a) Total condensate and reservoir particle number.  (b) Condensate and reservoir density at the spot center. (c) Flux of the runaway polaritons. In the linear stage, the condensate grows, depleting the reservoir. The main loss mechanism is due to the flux of runaway polaritons. In the self-trapping regime, in the vicinity of the pumping spot center, the condensate is dense and narrow, whereas in the reservoir there is a ``burned hole.'' Polaritons are strongly trapped, and the flux reverses its sign. Finally, at the blast stage, the condensate leaves the trap as a circular wave.     }
    \label{FigNumSM}
\end{figure}

To resolve the blast-stage dynamics beyond the weakly nonlinear regime, we integrated the full 2D ddGPE~\eqref{ddGPE1} directly on a polar grid using a GPU-accelerated pseudospectral scheme. A polar grid was used to avoid Cartesian symmetry breaking. The kinetic operator is diagonalized via a discrete Fourier--Bessel (quasi-discrete Hankel) transform: the angular dependence is expanded in a Fourier series in $\theta$, and each angular harmonic $m$ is propagated on its own natural radial grid, given by the zeros of $J_m$, where the Hankel transform is exact; cubic-spline collocation interpolates fields between this per-$m$ grid and the common physical radial grid. Reservoir, nonlinear, and absorbing-boundary terms are applied in real space via Strang (split-step) operator splitting. We used a domain of radius $r_{\max}=32$, $N_r = 1024$ radial collocation points, $N_\theta = 256$ angular points (harmonics $|m|\le 16$ retained), and time step $dt = 0.001$, evolved to $\tau = 100$. The explosion supernova-blast scenario shown in Fig.~\ref{FigAuto} corresponds to $r_0 = 2.5$, $p_0 = 20$, with $\varkappa=1$, $\gamma=1$, $g=4$, $\beta=0.1$. By increasing the pumping to $p_0=25$, we have stronger trapping and observe the ``tunneling'' scenario shown in the bottom-right picture of Fig.~\ref{FigAuto}.  
Details of particular simulations for the supernova-blast scenario are shown in Fig.~\ref{FigNumSM}.

\section{Details of experiment} 

The sample used in our experiments is a $3\lambda/2$ GaAs-based microcavity embedding GaAs$/$Al$_{0.15}$Ga$_{0.85}$As multiple quantum wells (MQWs), grown by metal-organic chemical vapor deposition (MOCVD) on a GaAs substrate. The sample was originally introduced in Ref.~\cite{choi2022realization} and is one of the samples used in Ref.~\cite{utesov2025universal} (referred to as KAIST 2 there). 
The cavity has the quality factor $Q \approx 8400$, with a vacuum Rabi splitting of $\hbar \Omega_\textrm{R}\approx 3.7$~meV. The lower polariton branch is centered at $E_\textrm{LP} \approx 1546$~meV.

The sample was placed in a 4 K cryostat and excited nonresonantly using a continuous-wave tunable laser (photon energy 1720 meV). The pump power was stabilized to within $1.5\%$ standard deviation throughout the course of the measurements. A spatial light modulator (SLM) was used to shape the excitation spot, allowing systematic control of the pump-spot diameter. Energy-resolved far-field photoluminescence spectra were acquired as a function of both excitation spot size and pump power to determine the different growth rates for each condensation regime.

\begin{figure}[t]
    \centering
    \includegraphics[width=0.66\linewidth]{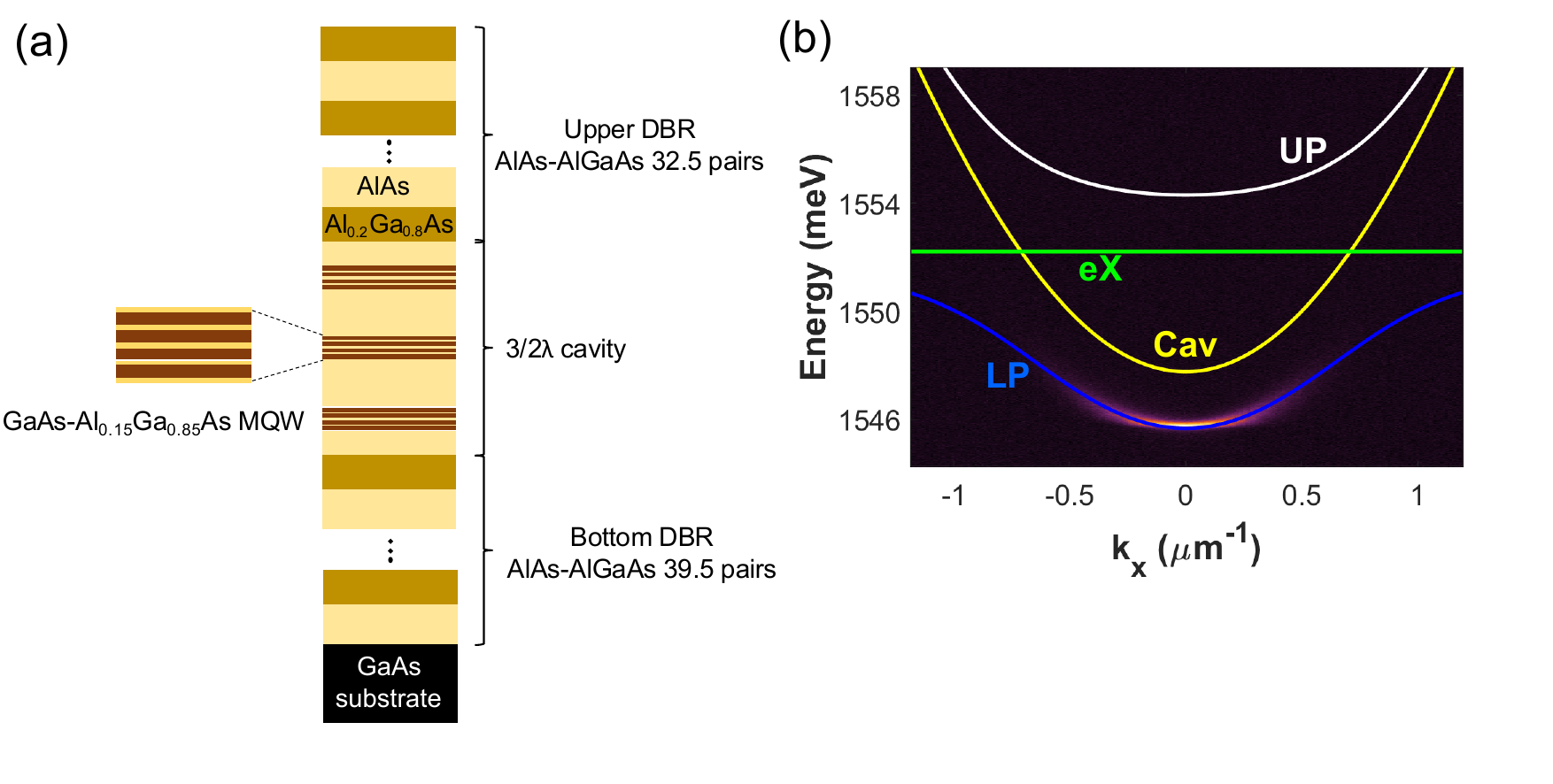}
    \caption{(a) Structure of multiple-quantum well (MQW) GaAs/AlGaAs sample used in our experiments. (b) Dispersion measured in the sample in the linear regime. LP -- low-polariton branch, UP -- upper-polariton branch, Cav -- cavity photon mode, eX -- exciton energy. }
    \label{FigExpSM}
\end{figure}

\begin{figure}[b]
  \centering
    \includegraphics[width=0.8\linewidth]{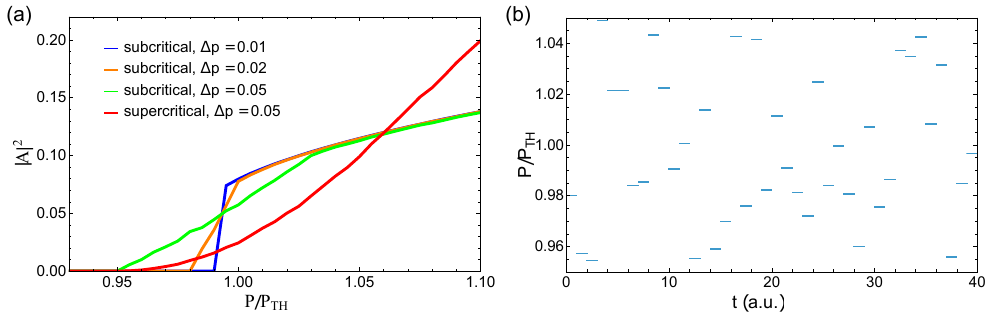}
      \caption{(a) Observable condensate spectral line intensity modeled for fluctuating power of the pumping laser. Green, orange, and green curves show the evolution of the hysteresis loop [see Fig.~2(d) of the Main Text] upon the fluctuation $\Delta p$ increase. (b) Example of time variation of the pumping power in the model for $\Delta p = 0.05$. Pumping power is taken from the box distribution with $2 \Delta p$ width and is fixed for a unit time; then the cycle repeats.  }
      \label{FigHystSupp}
\end{figure}

\section{Unstable laser and hysteresis}

We already pointed out that measuring the averaged signal only in the CW regime can lead to ambiguous results. The predicted hysteresis can be smeared out by laser power variations on relatively long time scales. For fits in Fig.~4 of the Main Text, we used a simplified minimal model presented in the End Matter. Here we discuss a more involved approach based on a Markov process formulation. In the limit when typical laser power variation substantially exceeds the hysteresis width, the results of both approaches are similar.

The algorithm is the following. We assume that the typical power variation time $T$ is much longer than the condensate formation time ($\sim 100$~ps for reasonable experimental parameters). Then, we randomly choose $p^\prime$ according to the box distribution in the $(p- \Delta p, p+ \Delta_p)$ range. Next, there are three cases
\begin{enumerate}
    \item for $p^\prime< p_\textrm{S}$, we ascribe $|A|^2=0$ to this step condensate density,
    \item for $p^\prime> p_\textrm{TH}$, we ascribe $|A(p^\prime)|^2 = |A_0|^2$ taken from Eqs.~(14) and~(15) of the Main Text to this step with the growth rate $\lambda^\prime \propto p^\prime-p_\textrm{TH}$,
    \item for $ p_\textrm{S}< p^\prime < p_\textrm{TH}$, the result depends on the condensate state on the previous step: if there was no condensate, it does not emerge, and if there was a condensate, it persists with the density recalculated for the current pumping power.
\end{enumerate}
Finally, we perform many steps and average the intensity over them.

The results of particular simulations are shown in Fig~\ref{FigHystSupp}(a), and the pumping power time variation is illustrated in Fig~\ref{FigHystSupp}(b). We took $p_\textrm{S}=0.98, \, p_\textrm{TH}=1$, and three values of $\Delta p = 0.01, 0.02, 0.05$. We also show the result for the standard supercritical regime with $\Delta p =0.05$. For each $p$, the results were averaged over 5000 steps. Note that this approach also predicts qualitatively different curve shapes for the subcritical regime with bistability and the supercritical Stuart-Landau regime. We used this property to identify the subcritical regime using our experimental data, see Fig.~4(b) of the Main Text.

\end{document}